\documentclass{ws-mpla}

\graphicspath{{../images/}}

\begin{document}

\markboth{J.~H.~WEBER}
{SINGLE QUARK ENTROPY AND THE POLYAKOV LOOP}

%%%%%%%%%%%%%%%%%%%%% Publisher's Area please ignore %%%%%%%%%%%%%%
\catchline{}{}{}{}{}
%%%%%%%%%%%%%%%%%%%%%%%%%%%%%%%%%%%%%%%%%%%%%%%%%%%%%%%%%%%%%%%%%%%

\title{SINGLE QUARK ENTROPY AND THE POLYAKOV LOOP
}

\author{\footnotesize JOHANNES HEINRICH WEBER}

\address{Physik Department T30f, Technische Universit\"at M\"unchen, James-Franck-Stra\ss e 1\\
Garching, Bayern 85748,
Germany\\
johannes.weber@tum.de}

\author{(TUMQCD collaboration)}

\maketitle

\pub{Received (Day Month Year)}{Revised (Day Month Year)}

\begin{abstract}
We study Quantum Chromodynamics (QCD) with 2+1 flavors with almost physical quark masses using the highly improved staggered quark action (HISQ). 
We calculate the Polyakov loop in a wide temperature range, obtain the free energy and the entropy of a single static quark and discuss the QCD crossover region in detail. 
We show that the entropy has a peak close to the chiral crossover and consider the consequences for the deconfinement aspects of the crossover phenomena. 
We study the renormalized Polyakov loop susceptibilities and place them into the context of the crossover. 
We also obtain a quantitative result for the onset of weak coupling behavior at high temperatures. 

\end{abstract}

\ccode{PACS Nos.: 12.38. Gc, 12.38.-t, 12.38.Bx, 12.38.Mh}

\section{Introduction}	

At sufficiently high temperatures, various gross properties of QCD matter are quite different from the respective properties at zero temperature. 
Namely, deconfinement of quarks and gluons, restoration of the isovector chiral symmetry and color screening are the key properties of the thermal QCD medium that distinguish it from the QCD vacuum at zero temperature (see \mbox{e.g.} Refs.\refcite{Bazavov:2015qsa,Meyer:2015wax} for some recent reviews).

The Polyakov loop is an operator that is sensitive to changes of the color screening properties of the medium.\cite{Polyakov:1980ca} 
It represents a static test charge in any particular representation of the gauge group. 
The bare Polyakov loop in the fundamental representation is defined with the lattice regularization as 
\begin{equation}
  L^{\rm{bare}}= \langle P \rangle, \qquad 
  P({\bf x})=\frac{1}{3} {\rm Tr}\, 
\prod_{x_0=0}^{N_{\tau}-1} U_0({\bf x},x_0),
 \label{eq:bare Polyakov loop}
\end{equation}
where $U_{\mu}(x=({\bf x},x_0))$ are the lattice link variables and $L^{\rm{bare}}$ is understood to be averaged over the lattice volume. 
$L^{\rm{bare}}$ has a linear UV divergence and needs renormalization.\cite{Polyakov:1980ca} 
The renormalized Polyakov loop is related to the free energy of a static quark, $F_Q$,\cite{McLerran:1981pb,Kaczmarek:2002mc} through 
\begin{equation}
 L^{\rm{ren}}= L^{\rm{bare}} \exp(-aC_Q) =\exp(-F_Q^{\rm{ren}}/T).   
\end{equation}
Since the renormalization of the Polyakov loop\footnote{
In the following, we usually mean the renormalized expectation value of the Polyakov loop in the fundamental representation when talking about the Polyakov loop without further specification.
} 
introduces a scheme dependence in terms of an additive constant in $F_Q^{\rm{ren}}$, only differences between $F_Q^{\rm{ren}}$ for different temperatures are observable quantities independent of the renormalization scheme.

In pure SU(N) gauge theories, the Polyakov loop is an order parameter of the transition. 
Below some transition temperature such gauge theories have a Z(N) center symmetry. 
Since the Polyakov loop in the fundamental representation is a center symmetry breaking field, its expectation value is strictly zero. 
The center symmetry is broken abruptly by the onset of color screening at the transition temperature. 
The fluctuations of the Polyakov loop diverge close to the transition temperature and the Polyakov loop acquires a real and positive expectation value. 
The discontinuous change is visible in both $L^{\rm bare}$ and $L^{\rm ren}$ . 
Hence, it can be used to define a deconfinement temperature that is independent of the renormalization scheme.\cite{Boyd:1996bx}

In QCD, this thermal transition is smoothed out by the sea quarks, and it is known that the QCD transition is a crossover.\cite{Aoki:2006we} 
The sea quarks explicitly break the center symmetry, giving preference to a small but positive expectation value of the Polyakov loop even at temperatures significantly below the crossover. 
This effect is understood in terms of the creation of a quark-antiquark ($Q\bar Q$) pair from the vacuum and the formation of static mesons. 
The fluctuations and the expectation value of the Polyakov loop are continuous functions and remain finite in QCD.

The Polyakov loop has been studied extensively\footnote{
A less incomplete list of references is found in Ref.~\refcite{Bazavov:2016uvm}. 
}  with weak-coupling methods,\cite{Berwein:2015ayt}
in SU(N) gauge theories with lattice methods,
\cite{Kaczmarek:2002mc,Gupta:2007ax}
and even in QCD with dynamical quarks close to or at the physical point.
\cite{Aoki:2006we,Bazavov:2016uvm,Aoki:2009sc,Bazavov:2013yv,Petreczky:2015yta} 
In the case of QCD, only the study in Ref.~\refcite{Bazavov:2016uvm} extends to sufficiently high temperatures for a comparison with weak-coupling approaches. 
Studies with larger quark masses found that the restoration of chiral symmetry and deconfinement happen at very similar temperatures,\cite{Karsch:2000kv,Petreczky:2004pz,Kaczmarek:2005ui} 
but the situation with quark masses close to the physical point might be different.

On the one hand the inflection point of $L^{\rm ren}$ suggests that deconfinement happens at significantly higher temperatures than the chiral crossover for realistic light quark masses,\cite{Aoki:2006we,Aoki:2009sc} 
while on the other hand a study of fluctuations of the Polyakov loop with an ad-hoc renormalization prescription suggests a similar temperature for deconfinement and chiral restoration.\cite{Lo:2013hla} 
Both observations can be reconciled if the scheme dependence of the Polyakov loop is actually taken into account.\cite{Bazavov:2016uvm}

This paper is organized as follows. 
In Sec.~\ref{sec:Lattice Setup}, we discuss our lattice setup. 
In Sec.~\ref{sec:Renormalization}, our use of three renormalization schemes and relations between them are covered. 
In Sec.~\ref{sec:Free Energy and Entropy}, we present continuum extrapolated results for the free energy and the entropy of a static quark and discuss the implications for the crossover and for the onset of weak-coupling behavior. 
In Sec.~\ref{sec:Polyakov loop susceptibilities}, we extract information on the crossover also from fluctuations of the Polyakov loop treated with gradient flow as the renormalization procedure. 
We finally conclude and summarize our results in Sec.~\ref{sec:Conclusions}.

\section{Lattice Setup}
\label{sec:Lattice Setup}

We calculate the Polyakov loop in QCD with 2+1 flavors on $N_\sigma^3 \times N_\tau$ lattices using the highly improved staggered quark (HISQ) action and a tree-level Symanzik improved gauge action.\cite{Follana:2006rc} 
This combination has leading discretization errors at $\mathcal O(\alpha_s a^2,a^4)$. 
We use lattices with $N_\tau=4$, $6$, $8$, $10$ and $12$ and an aspect ratio of $N_\sigma/N_\tau=4$, which is large enough to ensure that finite volume effects are small. 
The strange quark mass is at the physical point and the average light quark masses are given by $m_l/m_s=1/20$. 
These are slightly larger than the average physical mass and correspond to a pion mass of $m_\pi=161\,{\rm MeV}$ in the continuum limit. 

We use gauge configurations generated by the HotQCD collaboration,\cite{Bazavov:2011nk,Bazavov:2014pvz} which were used in high temperature studies of quark number susceptibilities.\cite{Bazavov:2013uja,Ding:2015fca}
We have generated new gauge configurations at higher temperatures with $N_\tau=4$, $6$ and $8$ and at lower temperatures for $N_\tau=10$ and $12$,  
the former to compare with weak-coupling results and the latter to reduce of uncertainties in the crossover region.\cite{Bazavov:2016uvm} 

The gauge configurations have been generated with the rational hybrid Monte Carlo (RHMC) algorithm using the MILC code.\cite{Bazavov:2010ru}
The inverse gauge couplings are in the range of $\beta=10/g_0^2=5.9$ - $9.67$. 
We fix the lattice spacing $a$ using the $r_1$ scale and its parametrization as in Ref.~\refcite{Bazavov:2014pvz}.
Since the inverse temperature is defined in the lattice regularization through $aN_\tau=1/T$, we observe that these gauge couplings correspond to a temperature range of $116\,{\rm MeV}<T<5814\,{\rm MeV}$. 
As we have calculated the Polyakov loop after every time unit (TU) of molecular dynamics, the statistical samples are usually between 30000 and 60000 TUs for temperatures up to $T \sim 400\,{\rm MeV}$. 
Though the samples for higher temperatures are usually much smaller, this is more than compensated by the smaller statistical fluctuations at high temperatures. 
Details of the gauge ensembles are found in Ref.~\refcite{Bazavov:2016uvm}.

\begin{figure}[h]
\centerline{\psfig{file=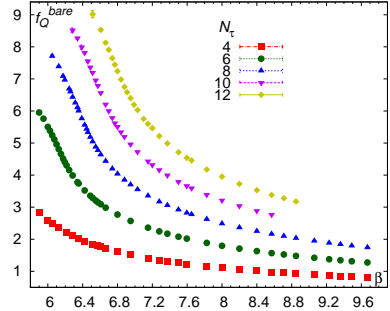,width=6cm}}
% \vspace*{8pt}
\caption{
The bare free energy of a static quark  
$ f_Q^{\rm bare}=F_Q^{\rm{bare}}/T= -\log L^{\rm{bare}} $ as function of the 
gauge coupling $\beta$ for different $N_{\tau}$ values.
\protect\label{fig: bare Fq}
}
\end{figure}

We show the bare free energy in units of the temperature, 
$f_Q^{\rm bare}=-F_Q^{\rm bare}/T = -\log L^{\rm bare}$, in Fig.~\ref{fig: bare Fq}. 
Cutoff effects at fixed temperature are varied through a simultaneous change of the coupling $\beta$ (\mbox{resp.} lattice spacing $a$) and the number of time slices $N_\tau$. 
Hence, the divergence of the bare Polyakov loop is evident, since the continuum limit would be approached going from the lower left towards the upper right corner in the figure. 
This divergence is removed in the process of renormalization.

\section{Renormalization}
\label{sec:Renormalization}

We renormalize the bare Polyakov loop using three different approaches. 
The first approach uses the static energy in the vacuum to obtain the renormalization constant in the fundamental representation and is suited for low to intermediate temperatures.  
The second approach uses relations between free energies at fixed temperature but different values of the cutoff and is suited for intermediate to high temperatures. 
The third is a fully independent approach -- the gradient flow. 
A renormalization constant $C_Q$ is added to the free energy in each approach,
\begin{equation}
 F_Q^{\rm ren}(T(\beta,N_\tau),N_\tau) = F_Q^{\rm bare}(\beta,N_\tau) + C_Q(\beta).
\end{equation}
Due to the leading discretization errors of the HISQ action, the renormalization constant can be parametrized as 
\begin{equation}
 C_Q(a) = \frac{b}{a} + c +\mathcal{O}(a^2), 
 \label{eq:renormalization constant}
\end{equation}
where the scheme-independent coefficient $b$ is determined by the requiring cancellation of the divergence. 
The scheme dependence is represented by the coefficient $c$ and the higher order terms. 
It is convenient to factor out the leading $a$ dependence by defining $c_Q=aC_Q$ and renormalize $f_Q^{\rm bare}$ through 
\begin{equation}
 f_Q^{\rm ren} = f_Q^{\rm bare} +N_\tau c_Q. 
 \label{eq:fqren}  
\end{equation}
In the following we usually omit the superscript ``$\rm ren$'' for renormalized quantities.

\subsection{Static quark-antiquark energy at $T=0$}
\label{sec:Static quark-antiquark energy at T=0}

The first approach relies on two relations, one between the Polyakov loop and the free energy of a static $Q\bar Q$ pair at large distances, and one between the free energy and the static energy of a static $Q\bar Q$ pair at short distances. 
It is known from weak-coupling calculations that the static energy\footnote{
In the following we always mean the static energy at zero temperature} 
and the free energy of a $Q\bar Q$ pair agree at short distances up to an additive term that is a trivial color factor.\cite{Berwein:2015ayt,Burnier:2009bk} 
Furthermore, the $Q\bar Q$ free energy assumes the asymptotic value $2F_Q$ for sufficiently large distances, where the two charges are uncorrelated due to color screening. 
Hence, $C_Q$ is given by half the renormalization constant of the static energy.

\begin{figure}[h]
\centerline{
\psfig{file=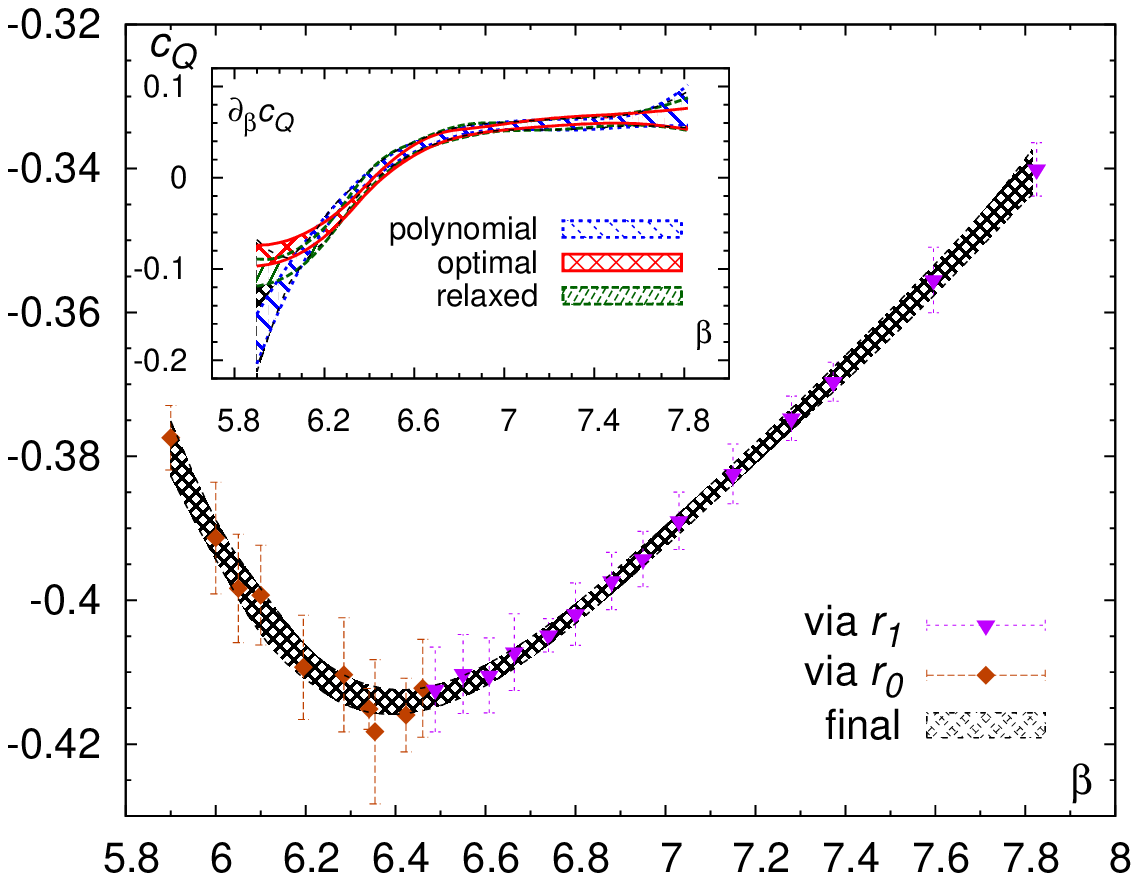,width=6cm}
\psfig{file=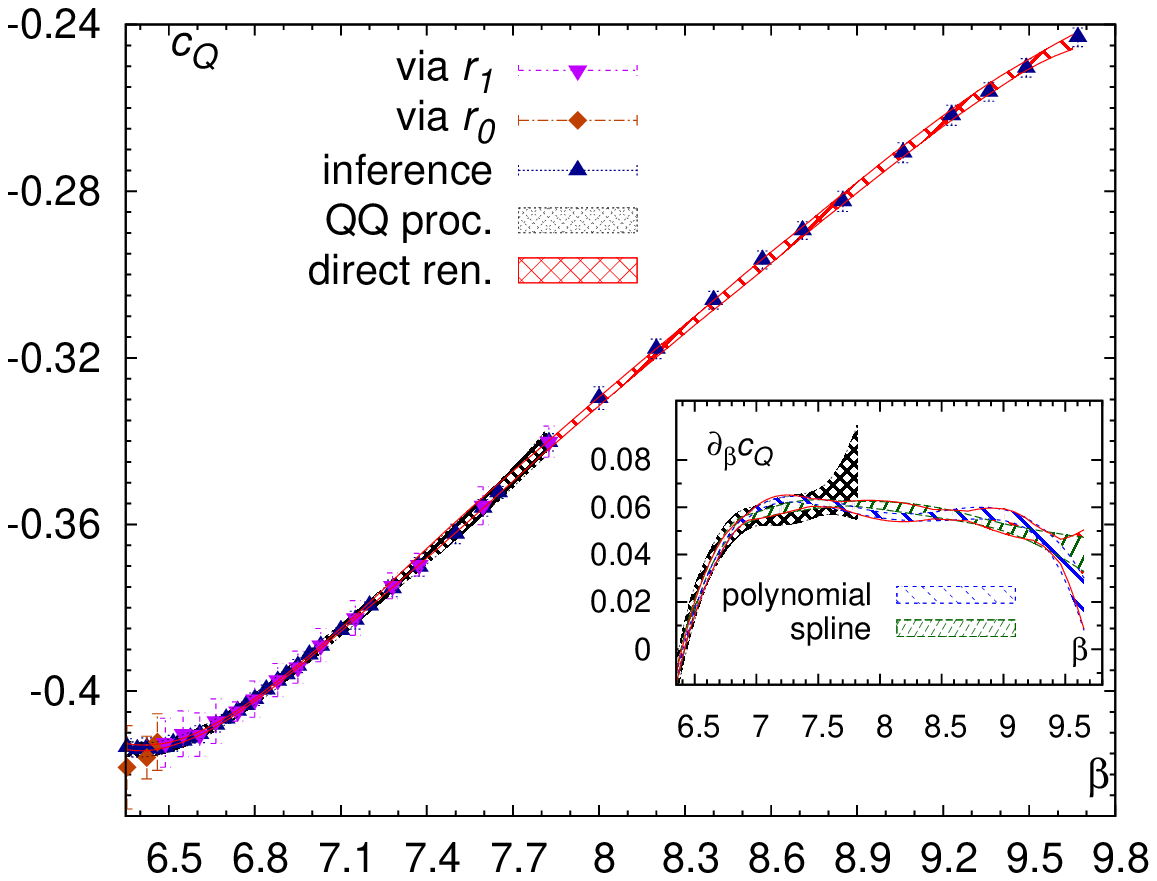,width=6cm}
}
% \vspace*{8pt}
\caption{
The renormalization constant $c_Q$ and its derivative obtained from the static energy (left) and from direct renormalization (right). 
Different interpolations are shown as $1\sigma$ bands. 
\protect\label{fig: renormalization constant cQ}
}
\end{figure}

$C_Q$ can be fixed in terms of the following procedure (so-called $Q\bar Q$ procedure) that is described in detail in Ref.~\refcite{Bazavov:2014pvz}. 
The static energy at finite cutoff is set to a prescribed value $0.954/r_0$ or $0.2065/r_1$ at the distances $r=r_0$ or $r_1$, where $r^2 d V/d r$ is equal to $1.65$ or $1.0$ respectively. 
The larger distance $r_0=0.4688(41)\,{\rm fm}$ is used for $\beta\leq 6.488$ (coarser lattices) and the smaller distance $r_1=0.3106(14)(8)(4)\,{\rm fm}$ is used for larger $\beta$ (finer lattices). 
After fixing the renormalization constant $C_Q$ with the static energy for $ 5.9 \leq \beta \leq 7.825$, we determine $c_Q$ for intermediate values of $\beta$ through an interpolation, which is discussed in detail in Ref.~\refcite{Bazavov:2016uvm}. 
This result is shown in the left panel of Fig. \ref{fig: renormalization constant cQ}.

\subsection{Direct renormalization}

The second approach relates the free energy at one value of $N_\tau$ to the free energy at another value of $N_\tau$.\cite{Gupta:2007ax} 
Once the renormalization constant $c_Q$ is known for one particular reference value $\beta^{\rm ref}$ (\mbox{e.g.} from the static energy), it can be inferred for different values of $\beta$, which correspond to the same temperature and different values of $N_\tau$. 
In order to invert the relation Eq. \eqref{eq:fqren}, one has to make sure that the free energy does not have any cutoff effects or one has to estimate these cutoff effects. 
Since the cutoff effects of the free energy have a significant temperature dependence for $T \lesssim 250\,{\rm MeV}$, they cannot be estimated well for low temperatures, where direct renormalization provides only a consistency check for the $Q\bar Q$ procedure, \mbox{cf.} the discussion in Ref.~\refcite{Bazavov:2016uvm}.  
Cutoff effects  are directly calculable from the difference 
\begin{equation}
 \Delta_{N_\tau,N_\tau^{\rm{ref}}}(T) 
 = f_Q(T(\beta,N_{\tau}),N_{\tau}) 
 - f_Q(T(\beta^{\rm ref},N_{\tau}^{\rm ref}),N_\tau^{\rm{ref}}). 
\end{equation}
Since the temperature dependence is rather mild for higher temperatures, we estimate an average of the cutoff effects, $\Delta_{N_\tau,N_\tau^{\rm{ref}}}^{\rm av}$, and infer the renormalization constant $c_Q$ for finer lattices as 
\begin{equation}
 c_Q(\beta) = \frac{1}{N_{\tau}} 
 \big[ N_{\tau}^{\rm{ref}} c_Q^{Q \bar Q}(\beta^{\rm{ref}}) 
 + \Delta_{N_\tau,N_\tau^{\rm{ref}}}^{\rm av} 
 + f_Q^{\rm{bare}}(\beta^{\rm{ref}},N_{\tau}^{\rm{ref}})
 - f_Q^{\rm{bare}}(\beta,N_{\tau}) \big].
\end{equation}
Using different pairs of $(N_\tau^{\rm ref},\beta^{\rm ref})$ for the same temperature, we check the consistency of the estimates of cutoff effects and iteratively extend the renormalization constant $c_Q$ to $\beta=9.67$. 
This extended result is shown in the right panel of Fig.~\ref{fig: renormalization constant cQ}. 
The technical subtleties of this procedure are covered in great details in Ref.~\refcite{Bazavov:2016uvm}.

\subsection{Renormalization with gradient flow}

The gradient flow is introduced as a device to remove short distance divergences in lattice observables.\cite{Luscher:2010iy,Luscher:2011bx} 
It is defined by a diffusion-type differential equation in an artificial fifth dimension $t$ -- the flow time -- which realizes a smoothing of UV fluctuations. 
The flow equation reads 
\begin{equation}
  \frac{d V_\mu(x,t)}{d t} = -g_0^2{\partial_{x,\mu} S[V]} V_\mu(x,t),
\end{equation}
where $g_0^2=10/\beta$ is the bare lattice gauge coupling, $\partial_{x,\mu}$ is a link differential operator as defined in Ref.~\refcite{Luscher:2010iy} and $S[V]$ is the lattice gauge action. 
The link variables at finite flow time are $V(x,t)$ with the initial condition $V_\mu(x,0)=U_\mu(x)$. 
Gradient flow finds much use for scale setting at zero temperature,\cite{Borsanyi:2012zs,Bazavov:2015yea} and it has been used in calculations of the Equation of State.\cite{Asakawa:2013laa} 
It was also used for calculating the renormalized Polyakov loop, which is obtained directly via Eq. \eqref{eq:bare Polyakov loop} using gauge links at finite flow time.\cite{Petreczky:2015yta} 
Hence, $C_Q$ does not appear explicitly in this scheme and the Polyakov loop is renormalized for arbitrary representations, such that the gradient flow is suitable for defining renormalized Polyakov loop susceptibilities. 
$L^{\rm ren}$ at finite flow time agrees up to a trivial scheme dependence in the continuum limit with the conventionally renormalized Polyakov loop for up to $T \sim 400\,{\rm MeV}$,\cite{Petreczky:2015yta} if the smoothing range $f=\sqrt{8t}$ induced by the gradient flows satisfies 
\begin{equation}
 a \ll f \ll aN_\tau =1/T.   
 \label{eq:flow condition}
\end{equation}
Using $L^{\rm ren}$ at finite flow time we find that satisfying Eq.~\eqref{eq:flow condition} turns out to be increasingly difficult for high temperatures ($T>400\,{\rm MeV}$). 
The flow time interval that reproduces the temperature dependence of $F_Q$ as observed in the direct renormalization scheme becomes very narrow and the flow times must be small. 
However, as these flow times are too small to remove the cutoff effects, we conclude that lattices with $N_\tau>12$ would be needed to use gradient flow for high temperatures.\cite{Bazavov:2016uvm}

\section{Free Energy and Entropy}
\label{sec:Free Energy and Entropy}

In the following we briefly discuss the extraction of the continuum limit for the free energy $F_Q$ and the entropy $S_Q$ and then discuss the consequences for the deconfinement aspects of the QCD crossover and for the high temperature regime. 
Finally we discuss the onset of the weak-coupling regime. 

\subsection{Continuum extrapolation of the free energy}

We extract the continuum limit of the free energy $F_Q$ with different procedures, whose properties are outlined hereafter and covered in detail in Ref.~\refcite{Bazavov:2016uvm}. 
Generally speaking, we split the fit ranges into intervals corresponding to low ($T \lesssim 200\,{\rm MeV}$), intermediate ($200\,{\rm MeV} \lesssim T \lesssim 400\,{\rm MeV}$) and high temperatures (up to a few GeV). 
The latter are only accessible in terms of the direct renormalization scheme. 

In so-called local fits, we interpolate $f_Q^{\rm bare}$ in $\beta$ for fixed $N_\tau$ using $\beta$ intervals that roughly separate the low temperature interval from the rest. 
The fits convergence well and smooth splines and polynomials yield consistent interpolations. 
Then we add ($N_\tau$ times) the renormalization constant and extrapolate for each temperature to the continuum limit.
In so-called global fits we simultaneously model the temperature and $N_\tau$ dependence of $f_Q^{\rm ren}$ in the form of 
\begin{equation}
 P(T,N_\tau) = P_0(T) + \frac{P_2(T)}{N_\tau^2} + \frac{P_4(T)}{N_\tau^4},
 \label{eq:continuum extrapolation} 
\end{equation}
where the $P_i(T)$ are polynomials in the temperature. 
The $N_\tau$ dependence is modeled according to the leading discretization errors of the HISQ action.  
For low temperatures $N_\tau=6$ data is not in the $1/N_\tau^2$ scaling regime. 
Thus, we omit $N_\tau=6$ data and $1/N_\tau^4$ term in the continuum extrapolation for low temperatures.

\begin{figure}[h]
\centerline{\psfig{file=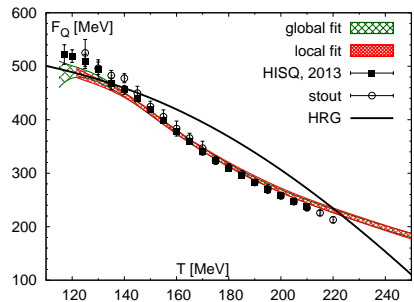,width=6cm}}
% \vspace*{8pt}
\caption{
The free energy of a static quark $ F_Q$ as function of the temperature in the vicinity of the crossover. 
The hadron resonnace gas model (HRG) is consistent with $F_Q$ up to $T \sim 135\,{\rm MeV}$. 
\protect\label{fig: Fq low}
}
\end{figure}

We estimate systematic uncertainties from the variation between different global and local fits, find them to be generally covered by the statistical uncertainties and eventually add these estimates in quadrature. 
The continuum limit from both local and global fits is shown in Fig.~\ref{fig: Fq low} together with older results and a calculation in the hadron resonance gas (HRG) model. 
The HRG is unsuitable as a description of the free energy already at $T \sim 140\,{\rm MeV}$. 
The new HISQ result supersedes the older HISQ result,\cite{Bazavov:2013yv} and it is reasonably close to the stout result from Ref.~\refcite{Borsanyi:2010bp}, which has slightly different light quark masses. 
The free energy does not show any pronounced features that prominently indicate the onset of deconfinement.

\subsection{Entropy}

In contrast to the featureless free energy, $F_Q$, the entropy of a static quark, $S_Q$, has well-pronounced features close to the crossover. 
The entropy is given through a temperature derivative of $F_Q$, 
\begin{equation}
 S_Q(T) = -\frac{\partial F_Q(T)}{\partial T}.
 \label{eq:entropy}
\end{equation}
This equality is still valid if the volume is not constant, since the static quark does not exert a non-zero pressure. 

The entropy has already been studied in pure gauge theory with lattice methods,\cite{Petreczky:2005bd,Kaczmarek:2005gi} and with holography approaches,\cite{Kharzeev:2014pha,Hashimoto:2014fha,Ewerz:2016zsx} where $S_Q$ is discontinuous at the phase transition. 
Studies in 2- or 3-flavor QCD with quark masses much larger than the physical masses found that $S_Q$ has a peak at the crossover temperature.\cite{Petreczky:2004pz,Kaczmarek:2005gi} 
This peak indicates the inflection point of $F_Q$. 
Hence, a continuum calculation of $S_Q$ at or close to the physical point is worthwhile as we illustrate hereafter. 

Using the local interpolations of $f_Q^{\rm bare}$ and $c_Q$, the entropy can be calculated at finite lattice cutoff as 
\begin{equation}
 S_Q 
 = \left(-1 + \frac{\partial }{\partial T}\right) \left(f_Q^{\rm bare} + N_\tau c_Q\right)
 = \left(-1+ T \frac{\partial \beta}{\partial T} \frac{\partial}{\partial \beta} \right) \left(f_Q^{\rm bare} + N_\tau c_Q\right),
 \label{eq:calculation of SQ}
\end{equation}
where the derivative $\partial \beta/\partial T$ is related to the non-perturbative running of the gauge coupling. 
This derivative can be traded for the non-perturbative beta function $R_\beta$ defined in Ref.~\refcite{Bazavov:2014pvz} through $R_\beta=T (\partial \beta/\partial T)$. 
Alternatively the entropy may be obtained directly from the previously discussed global fits for $f_Q$, which are polynomials in the temperature. 
Since the lattice spacing is related to the inverse temperature through $1/T=aN_\tau$ and $C_Q$ is given in terms of Eq. \eqref{eq:renormalization constant}, the scheme-dependent terms vanish the continuum limit of Eq. \eqref{eq:calculation of SQ},
\begin{equation}
  S_Q 
  = -\frac{d (F_Q^{\rm bare} \!+\! C_Q)}{d T}
  = -\frac{\partial \left(F_Q^{\rm bare} \!+\! \frac b a \right) }{\partial T} 
  + \frac{1}{T}\frac{\partial \left( c \!+\! \mathcal{O}(a^2) \right)}{\partial \log a}
  = -\frac{\partial F_Q}{\partial T} 
  + \mathcal{O}(a^2).
\end{equation}

\begin{figure}[h]
\centerline{
\psfig{file=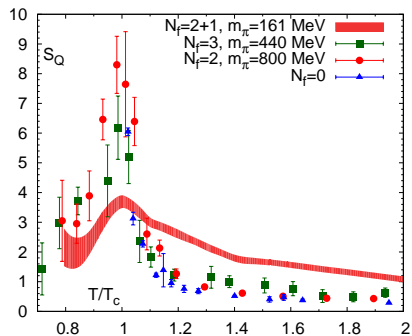,width=6cm}
}
% \vspace*{8pt}
\caption{
The entropy of a static quark $ S_Q$ as function of the temperature (in units of the pseudocritical temperature) in the vicinity of the crossover. 
\protect\label{fig: Sq cont}
}
\end{figure}

Hence, we find that the continuum limit of $S_Q$ is a scheme-independent observable and that the peak of $S_Q$ defines a scheme-independent temperature $T_S$, which is a characteristic of the deconfinement aspect of the crossover. 
We obtain $T_S=153^{+6.5}_{-5}\,{\rm MeV}$ in the continuum limit (statistical errors only) and find that $T_S$ varies in the range $150.5\,{\rm MeV} \leq T \leq 157\,{\rm MeV}$ for different fits, which defines our estimate of systematic uncertainties. 
We show our result for $S_Q$ in Fig. \ref{fig: Sq cont} together with older results at larger quark masses,~\cite{Petreczky:2004pz,Kaczmarek:2005gi}, which are not extrapolated to the continuum limit ($N_\tau=4$). 
The temperature axis has been rescaled by the corresponding lattice results for the pseudocritical temperature, which are cutoff dependent for the older results. 
The height of the peak of $S_Q$ is much reduced in the continuum result.

\subsection{Implications for the crossover}

\begin{figure}[h]
\centerline{
\psfig{file=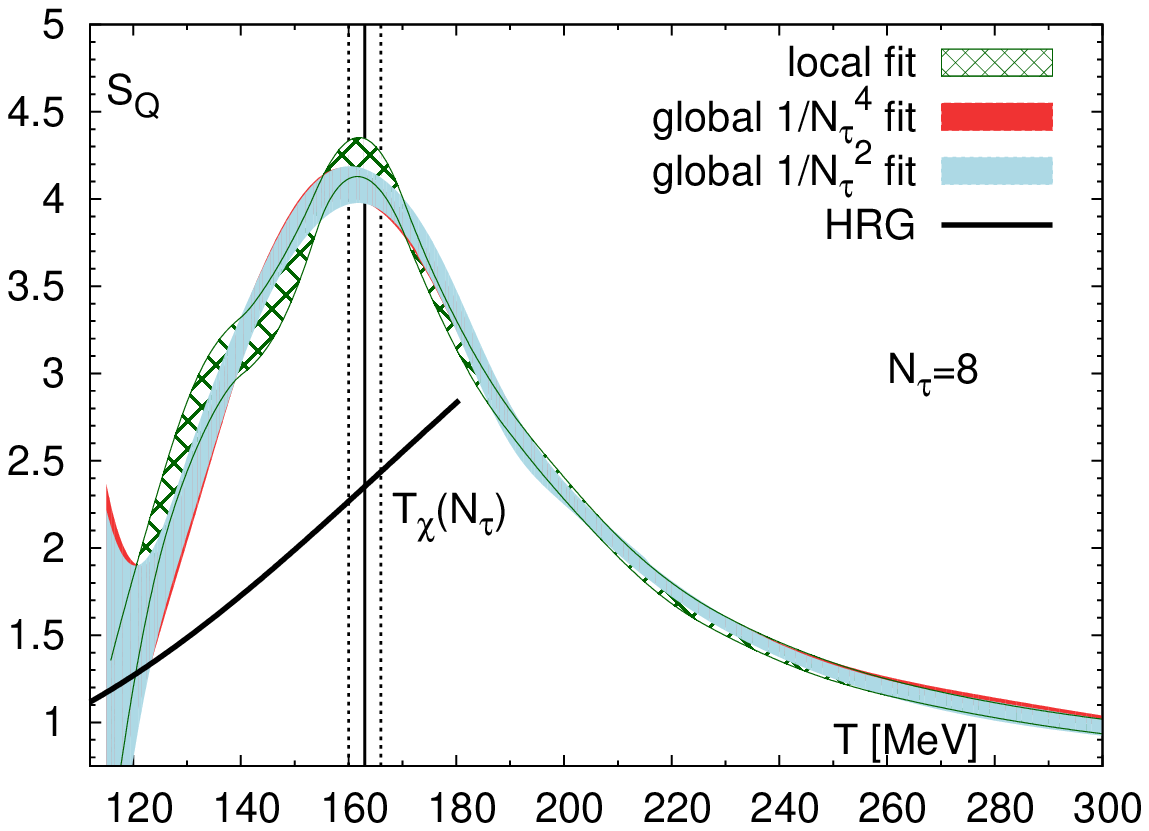 ,width=6cm}
\psfig{file=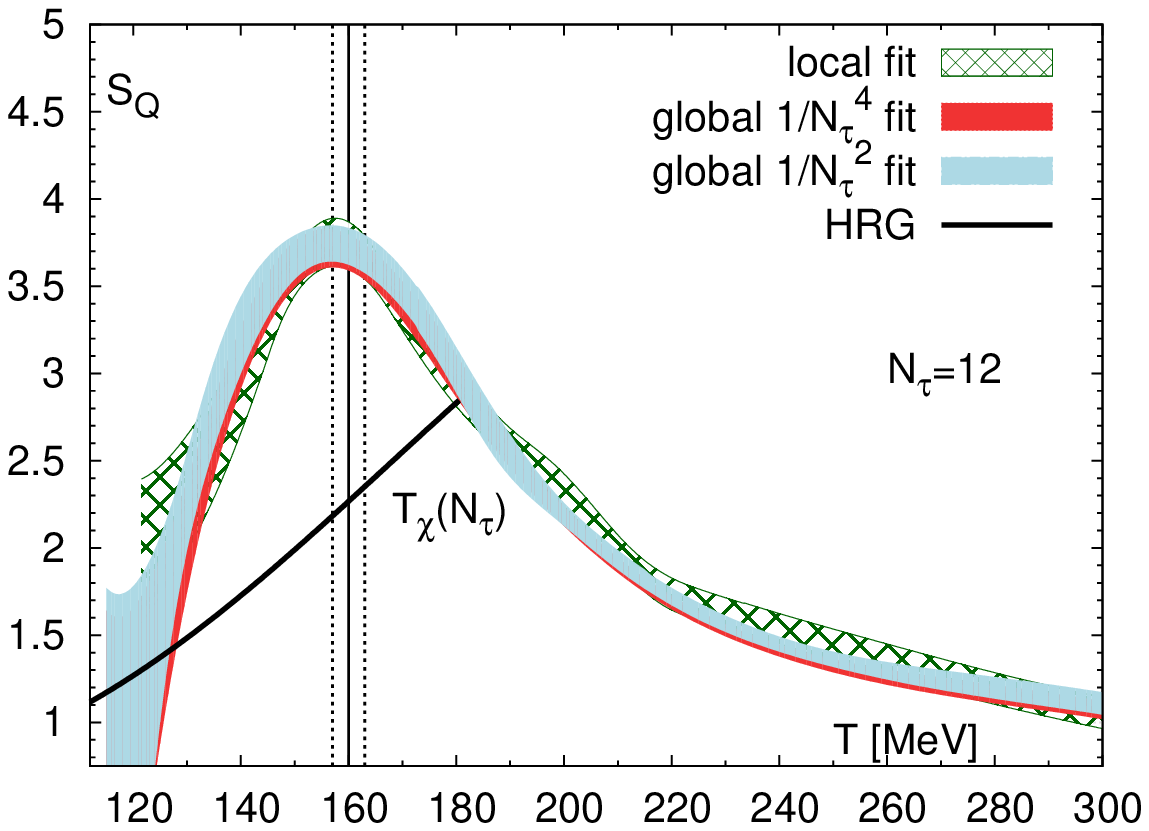,width=6cm}
}
% \vspace*{8pt}
\caption{
 $ S_Q$ at finite cutoff (left: $N_\tau=8$, right: $N_\tau=12$) as function of the temperature. 
 The peak position $T_S(N_\tau)$ has a similar cutoff dependence as the chiral crossover temperature $T_\chi(N_\tau)$ (shown as vertical bands). 
 The HRG model fails to describe the lattice result for $T > 125\,{\rm MeV}$. 
\protect\label{fig: Sq Ntau}
}
\end{figure}

We show $S_Q$ for two values of $N_\tau$ in Fig.~\ref{fig: Sq Ntau}. 
When approaching the continuum limit, we observe a fairly weak reduction of the peak height. 
This is definitely not enough to explain the smaller peak height of the continuum result in Fig.~\ref{fig: Sq cont} compared to the older results at finite cutoff and larger quark masses. 
This suggests that the lower quark mass might be the main reason for the diminished height of our peak.

We see that the HRG model for the entropy clearly breaks down at temperatures well below the actual peak. 
In such an HRG model, the Polyakov loop is given in the hadronic phase in terms of a sum over all meson and baryon states that include one static quark.\cite{Megias:2012kb} 
On the one hand, the renormalized energy of each static hadron amount to the sum of the dynamical quark masses and the respective binding energy in the HRG. 
Thus, the free energy of a single static quark in the hadronic phase should have a contribution linear in the dynamical quark masses. 
On the other hand, the free energy of a single static quark in the deconfined phase does not have the same linear quark mass dependence.
Thus, the HRG model suggests that the magnitude of the change of $F_Q$ during the crossover, which determines the height of the peak in $S_Q$, should be roughly linear in the light quark masses. 
For infinite quark masses $S_Q$ should diverge as seen in pure gauge theory. 
Though the trends in the lattice results match these expectations, it is not possible to draw quantitative conclusions given the size of errors of the older results with larger quark masses.

Fig.~\ref{fig: Sq Ntau} shows that the temperature $T_S$ and the chiral crossover temperature $T_\chi$\footnote{We use the value of $T_\chi(N_\tau)$ defined in terms of O(2) scaling fits at the same values of the quark masses, cf. Ref.~\refcite{Bazavov:2011nk} for a detailed discussion.} 
have a very similar cutoff dependence. 
This observation is consistent with the na\"ive idea that chiral restoration and deconfinement happen concurrently. 
However, if one defines the deconfinement temperature as the inflection point of the Polyakov loop as it is usually done, \mbox{e.g.} Refs.~\refcite{Aoki:2006we,Aoki:2009sc}, one finds $T_L=171(3)(4)\,{\rm MeV}$ or $T_L=170(4)(3)\,{\rm MeV}$. 
The observation that $T_L$ is consistently about $15$-$25$ MeV higher than the chiral crossover temperature is then considered as evidence for a fairly large width of the QCD crossover. 
However, this point of view opens up questions regarding the status of QCD matter for intermediate temperatures.

\begin{figure}[h]
\centerline{
\psfig{file=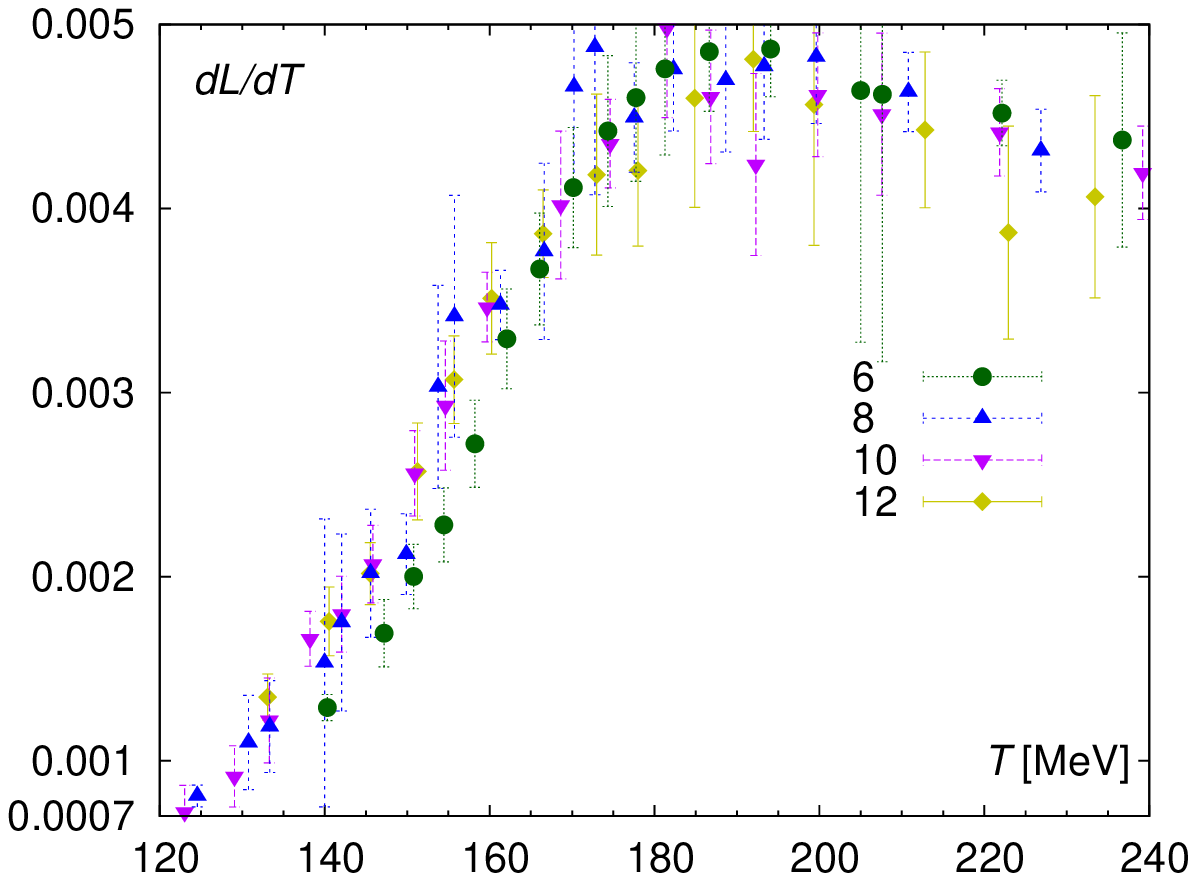 ,width=6cm}
\psfig{file=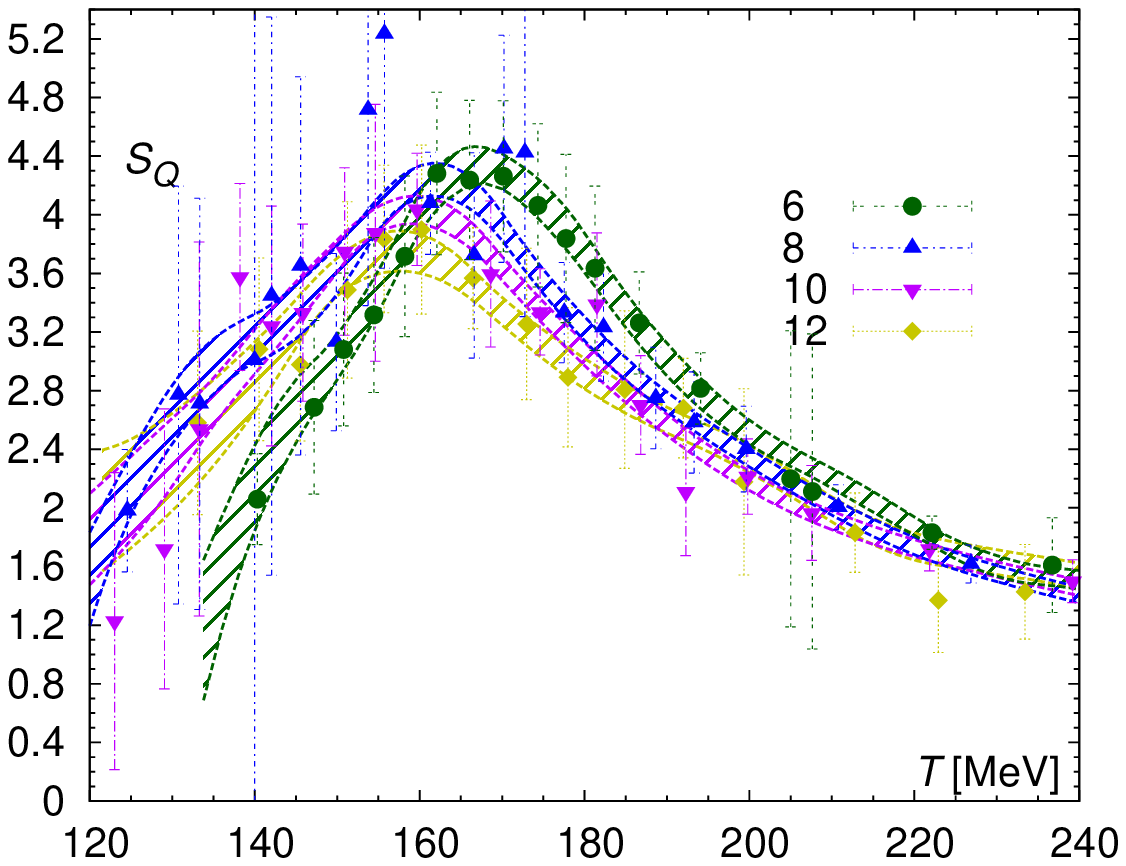,width=6cm}
}
% \vspace*{8pt}
\caption{
 Both $dL/dT$ (left) and $S_Q$ (right) are probes of the deconfinement of QCD matter.
\protect\label{fig: TL vs TS}
}
\end{figure}

In order to understand this issue, we first calculate the inflection point of the Polyakov loop from our data. 
We show the temperature derivative of the Polyakov loop in the left panel of Fig. \ref{fig: TL vs TS}, which has a peak between $180$ and $200$ MeV for all four considered values of $N_\tau$\footnote{
As the ordering of $dL/dT$ for different $N_\tau$ is not clear from the present data, we do not extrapolate to the continuum limit.  
A better result would have required generating more gauge ensembles. 
}. 
Thus, we find a value of $T_L$ that is significantly higher than $T_S$ (even at the same $N_\tau$), which supports the findings of Refs.~\refcite{Aoki:2006we,Aoki:2009sc}. 
However, we find that $L$ and therefore $T_L$ are scheme-dependent quantities and their temperature dependence is dominated by regular instead of singular terms. 
We express the relation defining $T_L$ in terms of $F_Q$, $S_Q$ and $d S_Q/dT$ and find 
\begin{equation}
 0 
 = \frac{1}{L} 
 \frac{\partial^2 L}{\partial T_L^2} 
 =
 \left[\frac{\partial f_Q}{\partial T_L}\right]^2 
 \!-\!
 \left[\frac{\partial^2 f_Q}{\partial T_L^2}\right] 
 =
 \frac{ F_Q^2 \!+\! 2[S_Q \!-\! 1]T_LF_Q}{T_L^4} 
 \!+\!
 \frac{S_Q^2 \!-\! 2S_Q \!+\! T_L\frac{\partial S_Q}{\partial T_L}}{T_L^2}.
 \label{eq:scheme dependence}
\end{equation}
Though the second term is singular and scheme-independent (in the continuum limit), the first term is regular and scheme-dependent through $F_Q^2$ and $F_Q$. 
A large change of $F_Q$ due to a change of the renormalization scheme cannot cancel (between $F_Q^2$ and $F_Q$) and must be compensated by a change in $T_L$ of a similar magnitude. 
In the light of this reasoning, we regard the scheme dependence of $T_L$ as a serious problem that may adversely affect conclusions on the width of the crossover.

\subsection{Weak-coupling limit}

\begin{figure}[h]
\centerline{
\psfig{file=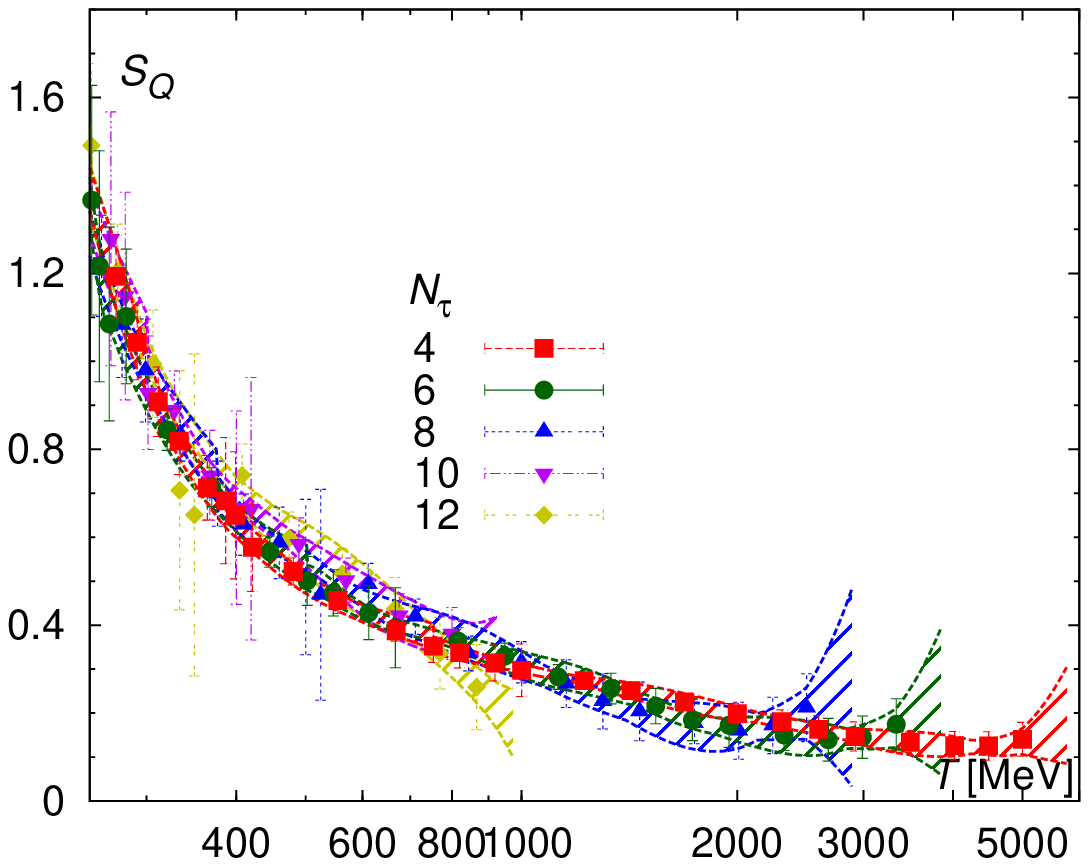 ,width=6cm}
\psfig{file=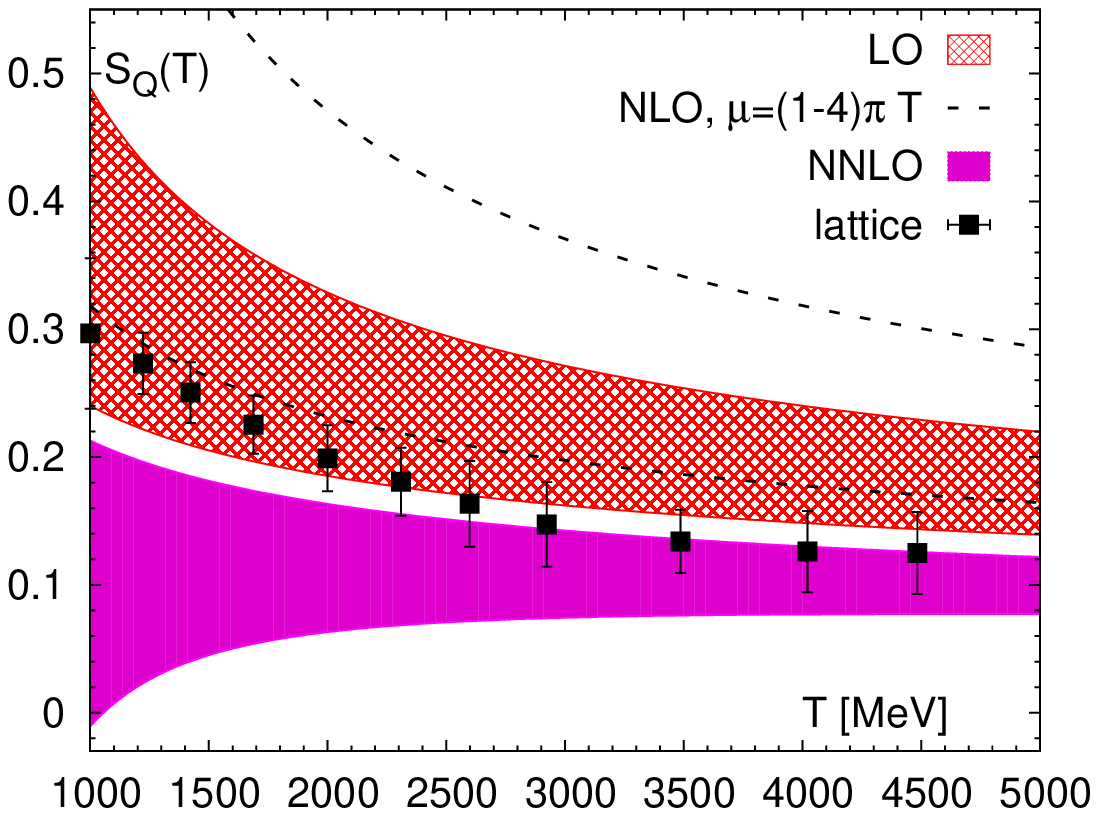,width=6cm}
}
% \vspace*{8pt}
\caption{
 $ S_Q$ at finite cutoff for high temperatures. 
 Cutoff effects are small (left) and consistency of the $N_\tau=4$ data with an NNLO weak-coupling calculation sets in for $T \sim 2.5\,{\rm GeV}$.
\protect\label{fig: SQ high}
}
\end{figure}

We compare our lattice results at high temperatures to weak-coupling calculations. 
Cutoff effects are very mild for temperatures above $T \sim 1\,{\rm GeV}$, as we show in the left panel of Fig.~\ref{fig: SQ high}. 
Since cutoff effects are smaller than the statistical uncertainties in the high temperature data of $S_Q$ for larger values of $N_\tau$, we do not attempt a continuum extrapolation here, but use results at finite values of the lattice cutoff.

At finite temperature, the weak-coupling expansion receives additional contributions form the Debye scale, $m_D \sim gT$, and the actual expansion parameter is $g$ instead of $\alpha_s=g^2/(4\pi)$. 
This expansion does not converge well, since $g \sim 1$ even at the scale of electroweak symmetry breaking. 
Moreover, the corrections from the next-to leading order account only for contributions suppressed by a single power of $g$, but fail to account for those corrections due to the next power in the usual $\alpha_s$ expansion.  This peculiar structure of the weak-coupling expansion at finite temperature is at the root of the large scale dependence of these weak-coupling results. 

The free energy has been calculated at next-to-next-to leading order (NNLO) in Ref.~\refcite{Berwein:2015ayt}, where the $\overline{MS}$ scheme was used. 
The lattice results, however, are defined in a scheme, where a fixed value is imposed on the static energy at some given distance, \mbox{cf.} Sec.~\ref{sec:Renormalization}. 
$F_Q$ in both schemes can be related by an additive matching constant that is explicitly calculable. 
Nevertheless, by considering $S_Q$, which is defined as a derivative of $F_Q$ in Eq.~\eqref{eq:entropy}, the matching is not necessary at all. 
Therefore, the direct comparison of lattice and weak-coupling results for $S_Q$ is straightforward.

We use lattice data with $N_\tau=4$ that extend up to temperatures as high as $T=5.8\,{\rm GeV}$, and show its comparison to the NNLO result in the right panel of Fig.~\ref{fig: SQ high}. 
For the NNLO result, we used 1-loop running of the coupling and a value $\Lambda_{\overline{MS}} = 315\,{\rm MeV}$, which was obtained from the static energy at zero temperature in Ref.~\refcite{Bazavov:2014soa}. 
The bands shown in the figure correspond to variation of the scale between $\mu=\pi T$ and $\mu=4\pi T$. 
The lattice result lies in between the leading order (LO) and NNLO results and approaches the latter from above. 
Numerically consistency is reached for $T \gtrsim 2.5\,{\rm GeV}$. 
This is qualitatively very similar to the results in pure gauge theory,\cite{Berwein:2015ayt} even though $S_Q$ is significantly larger in QCD. 
The size of the quark contribution is well-understood in terms of weak-coupling calculations. 

The onset of weak-coupling behavior in $S_Q$ at very high temperatures is in stark contrast to the case of quark number susceptibilities, where consistency with weak-coupling results is reached already for $T > 300\,{\rm MeV}$.\cite{Bazavov:2013uja,Ding:2015fca} 
This disparity is due to the fact that $S_Q$ is dominated by the static sector,\cite{Berwein:2015ayt} whereas the major contributions to the quark number susceptibilities originate in the non-static Matsubara modes. 
Let us note that the corrections to the leading order Debye mass are quite large in an extended temperature range.\cite{Karsch:1998tx}

\section{Polyakov loop susceptibilities}
\label{sec:Polyakov loop susceptibilities}

The Polyakov loop susceptibility is defined as 
\begin{equation}
 \chi = (VT^3) \left( \langle {|P|^2} \rangle -\langle {|P|} \rangle^2 \right)
\end{equation}
and finds much use for studying the deconfinement transition in pure gauge theories, where it has a sharp peak at the pseudocritical temperature.\cite{Datta:2015bzm} 
As the features of this peak are smoothened by the quarks in QCD, appropriate renormalization of $\chi$ is of crucial importance for the determination of the peak position. 
However, due to the mixing of representations, \mbox{e.g.} with the adjoint representation in
\begin{equation}
 | P_3 |^2 = |P_8| - 1, 
\end{equation}
and Casimir scaling violations, susceptibilities cannot be renormalized with the $Q\bar Q$ procedure as witnessed by the persistent UV divergences in Ref.~\refcite{Lo:2013hla}. 
Casimir scaling violations in QCD become small only for high temperatures ($T > 250\,{\rm MeV}$).\cite{Petreczky:2015yta}

Thus we renormalize the Polyakov loop susceptibilities using the gradient flow, \mbox{cf.} Sec.~\ref{sec:Renormalization}. 
We see a peak in $\chi$ that does not show a marked $N_\tau$ dependence and thus has only mild cutoff effects, but its central value has a strong flow time dependence. 
For the largest flow time considered, we find the peak at $T \simeq 200\,{\rm MeV}$, which is much closer to the scheme-dependent inflection point $T_L$ of the Polyakov loop than to the scheme-independent peak of the entropy at $T_S \sim 153\,{\rm MeV}$. 
This result indicates that $\chi$ is insensitive to the pseudocritical behavior in the crossover region and that its value is determined by scheme-dependent regular terms.

\begin{figure}[h]
\centerline{
\psfig{file=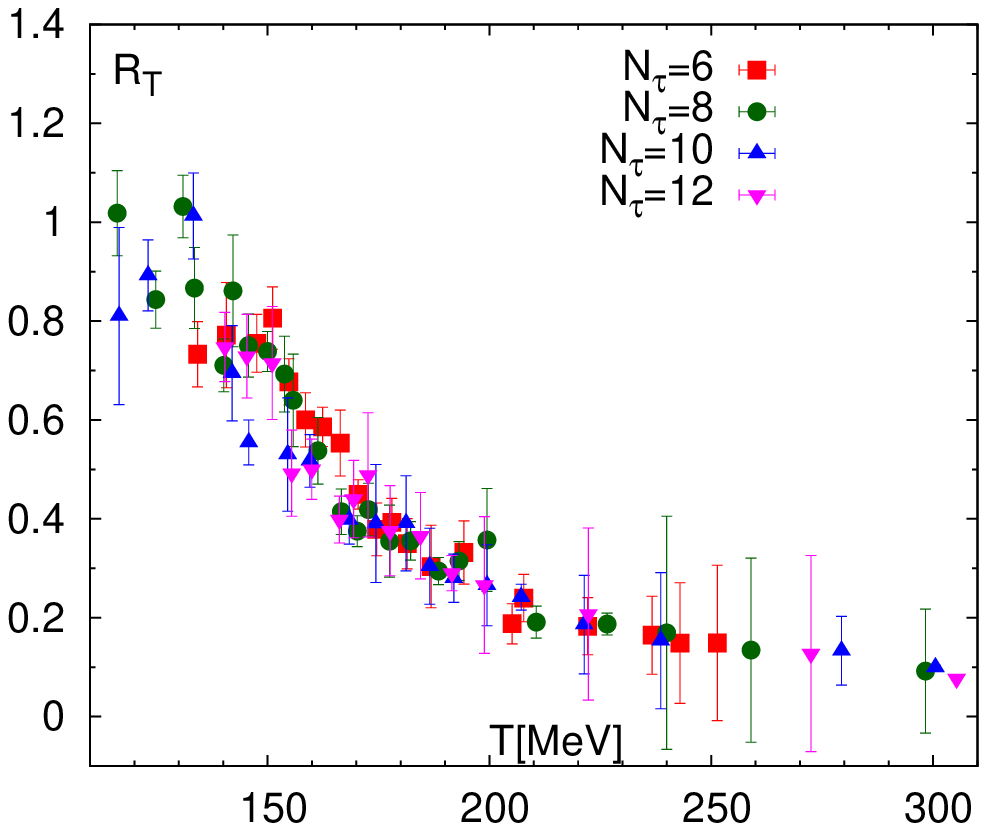,width=6cm}
\psfig{file=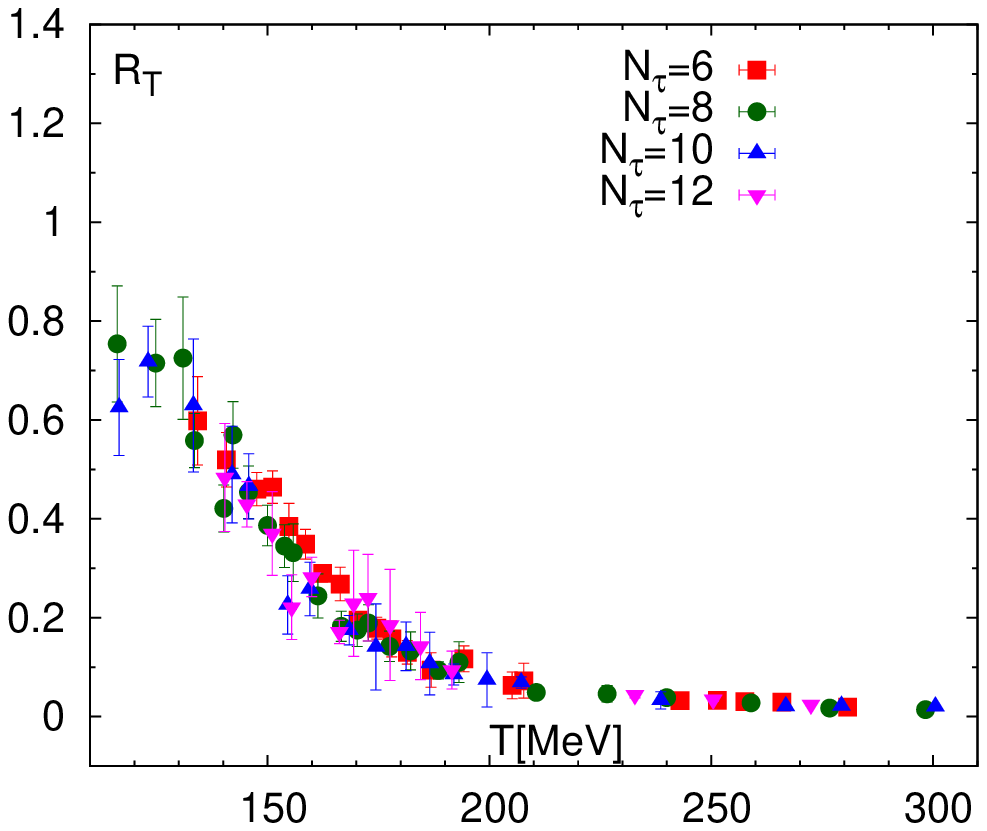,width=6cm}
}
% \vspace*{8pt}
\caption{
 $R_T=\chi_T/\chi_L$ for two different flow times, $f=1f_0$ (left) and $f=3f_0$ (right). 
\protect\label{fig: RT}
}
\end{figure}

Furthermore, we study the fluctuations of the real and imaginary parts of the Polyakov loop\footnote{
The expectation values of real and imaginary parts satisfy $\langle {{\rm Re}\,P} \rangle=\langle {P} \rangle$ and $\langle {{\rm im}\,P} \rangle=0$.
} individually, 
\begin{equation}
 \chi_L = (VT^3) \left( \langle { ({\rm Re}\,P)^2} \rangle -\langle {P} \rangle^2 \right), 
 \quad
 \chi_T = (VT^3) \langle { ({\rm Im}\,P)^2} \rangle 
\end{equation}
which we call longitudinal and transverse susceptibilities following Refs.~\refcite{Lo:2013hla,Lo:2013etb}. 
We find that the temperature dependence of $\chi_L$ is quite similar to the temperature dependence of $\chi$ and observe that $\chi_T$ is peaked near the crossover region. 
It has been argued in Refs.~\refcite{Lo:2013hla,Lo:2013etb} that ratios of Polyakov loop susceptibilities such as $R_A=\chi/\chi_L$ and $R_T=\chi_T/\chi_L$ probe deconfinement while being insensitive to the cutoff. 
Indeed, we do not observe a strong cutoff dependence in these ratios at finite flow time. 
For finite flow time, we see that $R_T$ exhibits crossover-like behavior in the vicinity of $T_S$, whereas $R_A$ is apparently insensitive to the crossover. 
Beyond some minimal value of the flow time both ratios show only mild flow time dependence.\cite{Bazavov:2016uvm}
$R_T$ for two flow times (in units of $f_0=0.2129\,{\rm fm}$) is shown in Fig.~\ref{fig: RT}.

\section{Conclusions}
\label{sec:Conclusions}

We have calculated the Polyakov loop in QCD with 2+1 flavors and realistic values of the sea quark masses using several lattice spacings over a wide temperature range. 
We have combined the $Q\bar Q$ procedure and the direct renormalization scheme to renormalize the Polyakov loop over the full temperature range. 
We have extracted continuum results for the free energy and the entropy and showed the scheme independence of the latter. 
Using gradient flow, we have also extracted renormalized Polyakov loop susceptibilities. 
We have argued that regular terms in the temperature dependence of the Polyakov loop and of the Polyakov loop susceptibility of its real part are introduced by their scheme dependence, such that these quantities may fail to be good probes for singular behavior in an arbitrarily chosen renormalization scheme. 
We have compared the $N_\tau=4$ lattice results for the entropy with a weak-coupling calculation at NNLO.

We find a peak of the entropy at $T_S=153^{+6.5}_{-5}\,{\rm MeV}$, which is a scheme-independent characteristic of the deconfinement aspect of the crossover. 
We show that $T_S$ has a similar cutoff dependence as the chiral crossover temperature, which supports the na\"ive idea that chiral symmetry restoration and deconfinement might be concurrent processes. 
We suggest that the singular behavior of the entropy can be interpreted in terms of the dissolution of static hadron states. 
The fluctuations of the imaginary part of the Polyakov loop seem to be sensitive to the crossover as well. 
Numerical consistency between lattice and weak-coupling results for the entropy is reached at very high temperatures, $T\gtrsim 2.5\,{\rm GeV}$. 
At significantly lower temperatures higher-order corrections are large and the weak coupling expansion is not reliable. 

\section*{Acknowledgments}

This work was supported by U.S. Department of Energy under 
Contract No. DE-SC0012704.
We acknowledge the support by the DFG Cluster of Excellence 
``Origin and Structure of the Universe'' (Universe cluster). 
The calculations have been carried out on Blue Gene/L computer of 
New York Center for computational Science in BNL,
at NERSC, at the Computational Center for 
Particle and Astrophysics (C2PAP) and
on SuperMUC at the Leibniz Supercomputer Center (LRZ). 
Usage of C2PAP and SuperMUC took place under the three Universe cluster 
grants ``Static Quark Correlators in lattice QCD at nonzero temperature'' 
for 2014, 2015 and 2016 (project ID pr83pu) and the LRZ grant 
``Properties of QCD at finite temperature'' for 2015 (project ID pr48le).
N.~Brambilla, A.~Vairo and J. H.~Weber acknowledge the support by the 
Universe cluster for the seed project ``Simulating the Hot Universe'', 
by the Bundesministerium f\"{u}r Bildung und Forschung (BMBF) under grant 
``Verbundprojekt 05P2015 - ALICE at High Rate (BMBF-FSP 202) GEM-TPC Upgrade 
and Field theory based investigations of ALICE physics'' under grant 
No. 05P15WOCA1 and by the
Kompetenznetzwerk f\"{u}r Wissenschaftliches H\"{o}chstleistungsrechnen 
in Bayern (KONWIHR) for the Multicore-Software-Initiative with the project 
``Production of gauge configurations at zero and nonzero temperature'' 
(KONWIHR-IV).

\end{document}